\documentclass[10pt,amsmath,amssymb,nofootinbib]{revtex4} 
\usepackage{dcolumn}
\usepackage{bm}

\newcommand{\be}{\begin{equation}}
\newcommand{\ee}{\end{equation}}
\newcommand{\bea}{\begin{eqnarray}}
\newcommand{\eea}{\end{eqnarray}}

\begin{document}
\title{Anatomy of a deformed symmetry: field quantization on curved momentum space}
\author{Michele Arzano}
\email{m.arzano@uu.nl}
\affiliation{Institute for Theoretical Physics,\\ Utrecht University,\\ Leuvenlaan 4, Utrecht 3584 CE, The Netherlands}
\begin{abstract}
\begin{center}
{\bf Abstract}
\end{center}
In certain scenarios of deformed relativistic symmetries relevant for non-commutative field theories particles exhibit a momentum space described by a non-abelian group manifold.  Starting with a formulation of phase space for such particles which allows for a generalization to include group valued momenta we discuss quantization of the corresponding field theory.  Focusing on the particular case of $\kappa$-deformed phase space we construct the one-particle Hilbert space and show how curvature in momentum space leads to an ambiguity in the quantization procedure reminiscent of the ambiguities one finds when quantizing fields in curved space-times.  The tools gathered in the discussion on quantization allow for a clear definition of the basic deformed field mode operators and two-point function for $\kappa$-quantum fields. 
\end{abstract}
\maketitle

\section{Introduction}
A characteristic feature of quantum field theory in curved space-time is that different observers, in general, do not agree on the particle content of {\it the same} quantum state of the field i.e. a ``natural" definition of particle does not exist \cite{Davies:1984rk}.  This is ultimately due to the lack of a unique choice of notion of time which guides the distinction between positive and negative frequency/energy modes and thus of particle and antiparticle at the quantum level.  Such ambiguity is already present when the space is a maximally symmetric one even though in these cases the symmetries available offer some criteria to pick up a particular choice of vacuum state.  The familiar Poincar\'e invariant vacuum in Minkowski space and the Bunch-Davies vacuum in de Sitter space are well known examples of such states.\\
The main motivation for the present work was the observation that, perhaps not surprisingly, similar issues regarding ambiguities in the definition of frequency/energy arise in a quite different setting namely for certain classes of ``non-commutative" field theories in which usual commuting space-time coordinates are replaced by generators of a Lie algebra.  In such theories, as discussed in detail in the rest of the paper, momentum space will turn into a non-abelian Lie group and thus into a curved manifold.  A natural question is why should one be interested in studying field theories defined on a curved momentum space.   One motivation comes from lower dimensional physics.  As first pointed out by 't Hooft \cite{'tHooft:1996uc} the momentum of a particle coupled to three-dimensional gravity as a conical defect is given by an angle leading to Lie algebra valued particle coordinates (see \cite{Matschull:1997du} and references therein for an extended discussion).  In higher dimension we encounter two more contexts in which field theories with curved momentum space play a major role.  On one side field theories defined on group manifolds are very useful tools in non-perturbative quantum gravity where they provide a way of generating  amplitudes for spin-foam models (see e.g. \cite{Oriti:2009wn}).  On the other hand certain models of non-commutative field theories are associated with momentum spaces given by homogeneous spaces other than the usual $\mathbb{R}^{3,1}$.  In these cases the curvature in momentum space introduces an energy scale which is {\it invariant} under the action of deformed relativistic symmetry generators \cite{Amelino-Camelia:2001fd, Arzano:2007ef, Arzano:2007gr, KowalskiGlikman:2003we, Freidel:2007yu,  Joung:2008mr, Meljanac:2010ps, Battisti:2010sr, Girelli:2009ii, Girelli:2010wi}.\\
Since the operational interpretation of non-commuting space-time coordinates is not immediate the starting point of our discussion will be a 
``symmetry based" description of the phase space of a relativistic particle alternative to the usual formulation in terms of cotangent bundle of a configuration space.  We will describe how this picture of a classical phase space naturally leads to the definition of a quantum one-particle Hilbert space.  However the crucial step that permits the distinction between particle and antiparticle states i.e. positive and negative energy states requires the introduction of a complex structure ``by hand".  This will be discussed in detail in Section III where we also recall how the arbitrariness of this choice is at the root of the ambiguity one encounters in the choice of vacuum state in curved space-times.  In Section IV we introduce the notion of ``curved" momentum space at the level of phase space focusing on a four dimensional model based on the $\kappa$-deformed Poincar\'e algebra where momentum space is embedded in a Lie group described by a sub-manifold of de Sitter space.  The structure of the momentum space group manifold is described in more in detail in the beginning of Section V as a preparation for the following discussion on the one-particle quantization from the deformed phase space and the related ambiguities.  In Section VI we provide a practical construction of the one-particle Hilbert space and field operators obtaining an explicit form of the two-point function and discussing the behaviour of quantum fluctuations of deformed field modes.  We conclude, in Section VII, with a summary of the results and a brief discussion.

\section{From particles to fields}
\subsection{Classical relativistic particle: phase space and symmetries}
In classical mechanics one has two equivalent ways of describing the phase space of a free relativistic particle. The usual approach is simply to take as the configuration space the ÒrangeÓ space of the coordinates of a particle (Minkowski space, $\mathbb{R}^{3,1}$) and define the (unreduced) phase space as the cotangent bundle of such configuration space.  The physical phase space will be given by a six-dimensional sub-manifold of the unreduced phase space whose coordinates parametrize geodesics in Minkowski space.  From an abstract mathematical point of view such phase space consists of a {\it symplectic manifold} $(\mathcal{M},\,\Omega)$, with $\mathcal{M}$ the cotangent bundle of the configuration space equipped with a closed non-degenerate two-form $\Omega$ (for more details see e.g. \cite{AbrMars}).\\
For a classical mechanical system which admits a continuous group of symmetries $G$ the phase space can be alternatively described by a group theoretic construction known {\it co-adjoint orbit method} \cite{Kir} which emphasizes the deep relation between $\mathcal{M}$ and $G$.  In this case the phase space can be constructed starting from the algebra $\mathfrak{g}^*$ dual to the Lie algebra $\mathfrak{g}$ of the symmetry group $G$.  Since the symmetry group $G$ has a natural {\it co-adjoint} action on $\mathfrak{g}^*$ the phase space manifold $\mathcal{M}$ will be given by the orbit $\mathcal{O}_Y$ of the co-adjoint action of $G$ on an element $Y\in \mathfrak{g}^*$.  The symplectic structure on $\mathcal{O}_Y$ will be induced by the natural symplectic structure on the dual algebra $\mathfrak{g}^*$.  The latter is defined as follows.  Take an element  $Y\in \mathfrak{g}^*$, since $\mathfrak{g}^*$ is a vector space the tangent space $T_Y\mathfrak{g}^*\simeq \mathfrak{g}^*$.  If we take a smooth function on the dual algebra $f\in C^{\infty}(\mathfrak{g}^*)$ then the differential $(df)_Y: T_Y\mathfrak{g}^* \rightarrow \mathbb{R}$ i.e. $(df)_Y$ can be seen as an element of the Lie algebra $\mathfrak{g}$ since $(df)_Y\in ( \mathfrak{g}^*)^*\simeq  \mathfrak{g}$.  The Poisson bracket on $C^{\infty}(\mathfrak{g}^*)$ is then given in terms of the commutators of $\mathfrak{g}$ by
\begin{equation}\label{LiePoiss}
\{f,g\}(Y)\equiv \langle Y, [(df)_Y, (dg)_Y]\rangle \,,
\end{equation}
where we used the natural pairing  $\langle Y, \xi\rangle$ of  $\mathfrak{g}$ and $ \mathfrak{g}^*$ as vector spaces.  The orbits $\mathcal{O}_Y$ of the co-adjoint action of $G$ on an element $Y\in \mathfrak{g}^*$ equipped with the symplectic structure above become symplectic manifolds which describe the phase spaces of $G$-symmetric mechanical systems.\\
In our specific context we are interested in the phase space of a relativistic point particle and thus we take the symmetry group $G$ to be the Poincar\'e group $ISO(3,1)= SO(3,1)\ltimes \mathbb{R}^{3,1}$.  In this case  $\mathfrak{g}^*=\mathfrak{iso}^*(3,1)\equiv \mathfrak{so}^*(3,1) \oplus (\mathbb{R}^{3,1})^*$ and the co-adjoint orbits $\mathcal{O}_{m,s}$ are given by level hyper-surfaces of the two Casimir functions $\mathcal{C}_1(p)$ and $\mathcal{C}_2(w)$ on $\mathfrak{iso}^*(3,1)$.  More specifically if we fix a set of co-ordinates $(p^0,\, p^i,\, j^i, k^i)$ on $\mathfrak{iso}^*(3,1)$ then we take $p=(p^0, p^i)$ and define the Pauli-Lubanski four vector $w=(w^0, w^i)$ by
\begin{equation}
 w^0={\bf p}\cdot{\bf j}\,,\,\,\,\,\,\,\,\, \vec{w}={\bf p}\times{\bf k}+p^0 {\bf j}\,.
\end{equation}
The mass and spin labels of the co-adjoint orbit $m$ and $s$ will be related to the fixed values of the functions $\mathcal{C}_1=p\cdot p$ and $\mathcal{C}_2=w\cdot w$.  Writing explicitly the Poisson structure on $\mathcal{O}_{m,s}$ for a specific choice of coordinate functions it can be seen \cite{Carinena:1989uw} that $\mathcal{O}_{m,s}\simeq \mathbb{R}^6\times S^2$ as a Poisson manifold i.e. a symplectic manifold describing the phase space of a relativistic spinning particle.  Notice here that the main advantage of the co-adjoint method approach is that it offers the most general formulation of a relativistic particleÕs phase space since encompasses the case of spinning particle which is normally not straightforward to describe in terms of the cotangent bundle on a configuration space \cite{Balachandran:1991zj}.\\
From here on we will focus on the phase space of a spinless relativistic particle.  In this case the Pauli-Lubanski vector vanishes identically and we denote the co-adjoint orbit by $\mathcal{O}_{m,0}$.  As mentioned above the dual algebra $\mathfrak{g}^*=\mathfrak{iso}^*(3,1)$ carries a natural Poincar\'e invariant Poisson structure directly related to the commutators of the Lie algebra $\mathfrak{g}=\mathfrak{iso}(3,1)$.  Indeed every $\xi\in \mathfrak{g}$ defines a linear co-ordinate function on $\mathfrak{g}^*$ given by $f_{\xi}$ such that $f_{\xi}(Y)=\langle Y, \xi \rangle$.  As we pointed out above for any function $f$ on $\mathfrak{g}^*$ the one-form $df$ can be seen as an element of the Lie algebra $\mathfrak{g}$.  In particular if we consider co-ordinate functions on $\mathfrak{g}^*$ associated with the generators of the Lie algebra $\xi_i$ then $df_{\xi_i}\equiv \xi_i$.  Denoting $h_i\equiv f_{\xi_i}$ it is easy to see that the Poisson brackets induced by the commutators of the Lie algebra $\mathfrak{g}$ will be given by
\begin{equation}
\{h_i, h_j\}=c^k_{ij} h_k\, ,
\end{equation}
where $c^k_{ij}$ are the structure constants of  $\mathfrak{g}$.  Starting from the coordinate functions $(p^0,\, p^i,\, j^i, k^i)$ on $\mathfrak{iso}^*(3,1)$ one can define a set of canonical co-ordinates on $\mathcal{O}_{m,0}$ using the spatial momentum coordinates $p^i$ and defining the position coordinates 
\begin{equation}
q^i=\frac{k^i}{p^0}\, ,
\end{equation}
with the coordinates satisfying the constraints $w^i =w^0 =0$ and $(p^0)^2-{\bf p}^{\,2}=m^2$.  Using the general formulae above it is easy to check that the canonical ``phase space" coordinates $\{q^i,p^i\}$ close the usual Poisson brackets
\begin{equation}
\{q_i, q_j\}=\{p_i, p_j\}=0\,,\,\,\,\,\,\,\,\,\,\{q_i, p_j\}=\delta_{ij}\, ,
\end{equation}
and thus $\mathcal{O}_{m,0}\simeq \mathbb{R}^6$ as expected.  Describing the phase space in terms of the co-adjoint orbit is in some way equivalent to consider a symplectic manifold whose natural coordinates are ``Poincar\'e momenta''.  Using co-adjoint orbits to describe phase space we  have a straightforward connection with the irreducible representations of $\mathfrak{iso}(3,1)$ since the latter are also labelled by the eigenvalues of the two invariant functions  $\mathcal{C}_1$ and $\mathcal{C}_2$.  We devote the rest of this section to such connection.\\

\subsection{Phase space of a classical field}
As a preparation for the discussion below it will be useful to make a short digression on the meaning of  ``positions" and ``momenta" when describing the phase space and symmetries of a relativistic particle.  Let us denote with $T$ the group of space-time translation.  For ordinary relativistic symmetries this is just $\mathbb{R}^{3,1}$ seen as a group under addition.  The Lie algebra $\mathfrak{t}$ of translation generators, as a tangent space to the identity element, can be identified with $\mathbb{R}^{3,1}$ as vector spaces.  The (trivial) Lie bracket on $\mathfrak{t}$ is induced by the addition law of the group $T\equiv\mathbb{R}^{3,1}$.  The dual group $T^*$ is, by definition, given by equivalence classes of unitary irreducible representations of $T$ and in the case $T\equiv\mathbb{R}^{3,1}$ elements of $T^*\equiv(\mathbb{R}^{3,1})^*$ are given by one-dimensional characters or in physics language `plane waves'.  When we write a plane wave like $e_{p}\in(\mathbb{R}^{3,1})^*$ we are simply saying that such element of  the dual group $(\mathbb{R}^{3,1})^*$ has coordinates given by the four-vector $p$.\\
One usually refers to 'positions' (as elements of the ambient space on which we build the (unreduced) configuration space) as given by coordinates on Minkowski space i.e. the translation group $T\equiv\mathbb{R}^{3,1}$.  Indeed, in the usual description, the unreduced phase space of a non-spinning relativistic particle is given by the cotangent bundle of the group of translations $T$ which is isomorphic \cite{AbrMars} to $T \times \mathfrak{t}^*$.  From this point of view 'momenta' are just coordinates on the dual Lie algebra $\mathfrak{t}^*$.  Let us point out that one also speaks of 'momenta' when referring the space-time translation generators i.e. a basis of the Lie algebra $\mathfrak{t}$.  In this case space-time `coordinates' correspond to the basis of generators the dual algebra $\mathfrak{t}^*$.  In ordinary relativistic theories we can refer to coordinates and momenta without specifying the objects we are referring to because $T$ and $\mathfrak{t}$ can be identified and so can their duals $T^*$ and $\mathfrak{t}^*$.  Notice how, instead, from the more general point of view of the co-adjoint orbit description of phase space it is only correct to say that the dual algebra $\mathfrak{iso}^*(3,1)$ provides the ambient space on which both position and momenta are defined.  As we will see in Section IV the distinction between $T$, $T^*$ and their respective Lie algebras will be crucial when momentum space becomes ``curved".  In that context a description of phase space in terms of co-adjoint orbits will provide a very clear characterization of the structures that lie at the basis of symmetry deformation.\\
Going back to our spinless relativistic particle, in the language of co-adjoint orbits its ``momentum space" will be given by the subspace $M_{m}\subset \mathcal{O}_{m,0}$ (the ``mass-shell") obtained by considering the restriction to the co-adjoint orbits of the abelian subalgebra $\mathfrak{t}^*\equiv(\mathbb{R}^{3,1})^*$ of $\mathfrak{g}^*=\mathfrak{iso}^*(3,1)$ dual to the algebra of translation generators.  Since for a ordinary relativistic particle in Minkowski space we can identify $\mathfrak{t}^*$ with $T^*$ the momentum space $M_{m}$ can be characterized in a {\it coordinate independent} way as a orbit of a character (``plane wave") under the action of the group $SO(3,1)$ (see \cite{Barut:1986dd}), i.e.
\begin{equation}
M_m \equiv\{\gamma e_{p}: e_{p} \in (\mathbb{R}^{3,1})^*,\, \gamma\in SO(3,1) \}\, ,
\end{equation}
which, keeping in mind the discussion above, can be described in terms of the co-ordinate functions on the dual algebra $\mathfrak{t}^*\equiv(\mathbb{R}^{3,1})^*$ by the two-sheeted hyperboloid $(p^0)^2-{\bf p}^{\,2}=m^2$.  
From its definition as a orbit of a symmetry group $M_m$ has a natural structure of a homogeneous space, indeed
\begin{equation}
M_m \simeq SO(3,1)/SO(3)\, ,
\end{equation}
with $SO(3)$ the ``isotropy" subgroup of $SO(3,1)$ which leaves invariant the point $(m,0,0,0)$.  Like any homogeneous space (under some additional assumptions see Barut pag 130) $M_m$ admits an invariant measure on its space of functions.  On the space of complex valued functions on the mass-shell $C^{\infty}(M_m )$ we can define the invariant measure $d\mu_m$ using the following trick \cite{Sternberg:1994tw}: one looks for the volume 3-form which satisfies
\begin{equation}
dV= d (\mathcal{C}_1(p))\wedge d\mu_m
\end{equation} 
where $dV$ is the ordinary volume 4-form on $\mathbb{R}^{3,1}$.  The invariant measure on $C^{\infty}(M_m )$ can be usefully written as a ``$\delta$-measure"
\begin{equation}
d\mu_m = d V \delta(\mathcal{C}_1(p))\, .
\end{equation} 
In the same spirit we can think of elements of $C^{\infty}( M_{m} )$ as distributions on $(\mathbb{R}^{3,1})^*$ given by
\begin{equation}
\tilde{\phi}(p)= \delta(\mathcal{C}_1(p)) \tilde{f}(p)
\end{equation}
with $\tilde{f}(p)\in C^{\infty}((\mathbb{R}^{3,1})^*)$.  A necessary and sufficient condition for a distribution to be of the form above is that
\begin{equation}
(\mathcal{C}_1(p)-m^2)\tilde{\phi}(p)=0\, .
\end{equation}
On the space of functions $C^{\infty}((\mathbb{R}^{3,1})^*)$ we can introduce a notion of Fourier transform which is just a special (trivial) case of the general Fourier transform of functions on a group (which will be useful later on)
\begin{equation}
f(\Lambda) = (d_{\Lambda})^{-1} \int_{G}d\mu(g)\, \tilde{f}(g)\, \hat{n}_{g}(\Lambda)
\end{equation}
where $\Lambda$ is an index of an irreducible representation of $G$, $d_{\Lambda}$ its dimension  and $\hat{n}_{g}(\Lambda)$ the character of such representation.  In our particular case for $\tilde{f}(p)\in C^{\infty}((\mathbb{R}^{3,1})^*)$ and $\tilde{\phi}(p)\in C^{\infty}( M_m )$ one has the familiar expressions 
\begin{equation}
f(x) = \int_{(\mathbb{R}^{3,1})^*}d\mu(p)\, \tilde{f}(p)\, e_p(x)\,,\,\,\,\,\,\,\,\,\,\,\,\,\phi(x) = \int_{(\mathbb{R}^{3,1})^*}d\mu(p)\, \delta(\mathcal{C}_1(p))\, \tilde{f}(p)\, e_p(x)\, .
\end{equation}
where $d\mu(p)=\frac{d^4 p}{(2\pi)^{3/2}}$ and $e_p(x)=\exp(-i p x )$.
Finally noting that under Fourier transform $\partial_i \phi(x)\rightarrow i p_i \tilde{\phi}(p)$  we have that 
\begin{equation}
(\mathcal{C}_1(p)-m^2)\tilde{\phi}(p)=0\,\Longleftrightarrow (\Box+m^2)\phi(x)=0\,,
\end{equation}
the Fourier transform maps functions on the mass shell hyperboloid into the space of solutions of the Klein-Gordon equation $\mathcal{S}$.
Notice that we also have $-\partial_i \phi^*(x)\rightarrow -i p_i \tilde{\phi}^*(p)$ and due to the quadratic nature of the equations above we make the identification $\tilde{\phi}^*(p)=\tilde{\phi}(-p)$ and $\phi^*(x)=\phi(x)$ i.e. the solutions of the Klein-Gordon equation are {\it real valued} functions.  The phase space of a classical field is then given by the symplectic manifold $(\mathcal{S},\omega)$ with symplectic structure provided by the antisymmetric bilinear form $\omega$ given by the Wronskian\footnote{In Minkowski space the integral is taken over a Cauchy surface $\Sigma_t$ at fixed time $t$
\begin{equation*}\label{qftinnprod}
\omega(\phi_1,\phi_2)= \int_{\Sigma_t}(\phi_2\dot{\phi}_1-\phi_1\dot{\phi}_2)\, d^3\vec{x}\, .
\end{equation*}}  associated to the Klein-Gordon equation
\begin{equation}\label{omega}
\omega(\phi_1,\phi_2)=\int_{\Sigma}(\phi_2\nabla_{\mu}\phi_1-\phi_1\nabla_{\mu}\phi_2)d\Sigma^{\mu}\, .
\end{equation}
This exhibits nicely the connection between phase space of a relativistic spinless point particle and the phase space of a classical scalar field.  Let us remark here that in Minkowski space (and in general on any globally hyperbolic space) the field's phase space is given by an equivalent description in terms of the space of initial data $\{\varphi, \pi\}$ on a given Cauchy surface $\Gamma_{\Sigma}$ with the symplectic form given by the restriction of $\omega$ above to such space.  In the next section we will discuss how a natural structure of inner product can be defined on the field's phase space and how this can be used to construct the ``one-particle" Hilbert space of the corresponding quantum field theory.

\section{Complex numbers and field quantization}
As we discussed above classical fields are {\it real} fields.  In classical field theory complex variables are often used as a computational tool with no physical meaning.  When we turn to the quantum setting however complex numbers become fundamental.  From the point of view of quantum observables the imaginary unit $i$ is introduced in order to turn differential operators into self-adjoint operators (e.g. momenta as generators of translations).  From the point of view of quantum states these are now rays of a {\it complex} Hilbert space.  Indeed, from a modern perspective, the very concept of quantization of a classical field amounts to the introduction of an appropriate {\it complex structure} $J$ on the classical phase space of the theory \cite{Bongaarts:1971cu, Ashtekar:1975zn, Panangaden:1979mi, Gibbons:1993iv}.\\
In the section above we discussed how the phase space of a classical field can be described by the space of solutions of the classical equations of motions $\mathcal{S}$.  This characterization of phase space will give an intuitive physical interpretation of the role of the complex structure since as we will see in more detail below, $J$ provides a direct sum decomposition of the {\it complexification} of $\mathcal{S}$, $\mathcal{S}^{\mathbb{C}}$ into ``positive and negative energy" subspaces which will represent, respectively, the ``one-particle" Hilbert space of the theory $\mathcal{H}$ and its complex conjugate $\bar{\mathcal{H}}$ once they are equipped with an appropriate inner product.  Of course the choice of $J$ is not unique but in certain specific cases it will be dictated by further physical inputs.  For example for a real scalar field in Minkowski space there exists a unique Poincar\'e invariant complex structure and it corresponds to the familiar textbook decomposition of the field in positive and negative frequency modes.  In more general space-times there will be no unique choice of $J$ and this is at the basis of the well known phenomenon of particle creation.  In this case different observers  will decompose the field according to different notion of positive and negative energy and will define different vacuum states for their quantum field.  From a more fundamental point of view such observers are just choosing different complex structures in representing the Hilbert space of their quantum field theory.\\
Let's try to be more concrete.  To introduce a {\it complex structure} on $\mathcal{S}$ amounts to define an automorphism $J:\mathcal{S}\rightarrow\mathcal{S}$ such that $J^2=-1$. 
As we mentioned above the introduction of $J$ corresponds to a choice of decomposition of $\mathcal{S}^{\mathbb{C}}$ in positive and negative energy subspaces.  Recall that the complexification $\mathcal{S}^{\mathbb{C}}$
of $\mathcal{S}$ is defined by
\begin{equation}
\mathcal{S}^{\mathbb{C}}\equiv\mathcal{S}\otimes\mathbb{C}\,.
\end{equation}
The complex linear extension of $J$ to $\mathcal{S}^{\mathbb{C}}$ is given by
\begin{equation}
J(\phi\otimes z)\equiv J(\phi)\otimes z\, .
\end{equation}
The introduction of $J$ gives rise to a natural decomposition of $\mathcal{S}^{\mathbb{C}}$ into two subspaces, $\mathcal{S}^{\mathbb{C}+}$ and $\mathcal{S}^{\mathbb{C}-}$ spanned, respectively, by the eigenvectors of $J$ with eigenvalues $\pm i$ i.e. $J(\phi^{\pm})=\pm i(\phi^{\pm})$.  We can define projectors $P^{\pm}:\mathcal{S}\rightarrow\mathcal{S}^{\mathbb{C}\pm}$
\begin{equation}
P^{\pm}\equiv \frac{1}{2}(1\mp iJ)\,,
\end{equation}
with
\begin{equation}
\mathcal{S}^{\mathbb{C}}=\mathcal{S}^{\mathbb{C}+}\oplus \mathcal{S}^{\mathbb{C}-}\,.
\end{equation}
The connection with positive and negative energy decomposition is now easily seen.  If the background space-time  admits a timelike and hypersurface orthogonal Killing vector field $\mathcal{L}_t$, i.e. it is {\it static}, one can decompose any real solution $\phi\in\mathcal{S}$ in normal modes (e.g. plane waves) of positive and negative energy components with respect to $\mathcal{L}_t$
\begin{equation}
\phi=\phi^+ +\phi^-\,.
\end{equation}
Then the map $J=-(-\mathcal{L}_t\mathcal{L}_t)^{-1/2}\mathcal{L}_t$ is such that
\begin{equation}
J\phi=i\phi^+ +(-i)\phi^-\,,\,\,\,\,\,\,\,\,\,\,\,\,P^{\pm}\phi= \phi^{\pm}
\end{equation}
i.e. $J$ is a complex structure on $\mathcal{S}$ and it provides a decomposition of $\mathcal{S}^{\mathbb{C}}$ in positive and negative energy subspaces.  Put the other way around {\it a decomposition of $\mathcal{S}^{\mathbb{C}}$ in positive and negative energy subspaces singles out a preferred complex structure $J$}.  Of course in order to obtain the ``one-particle" Hilbert space $\mathcal{H}$ from $\mathcal{S}^{\mathbb{C}+}$ we need to equip the latter with a positive definite inner product.  This can be constructed using $J$ itself and the natural symplectic structure (\ref{omega}) of the classical phase space under the further requirement that the complex structure be {\it compatible} with the symplectic structure $\omega$, namely 
\begin{equation}
\omega(J\phi_1,J\phi_2)=\omega(\phi_1,\phi_2)\, .
\end{equation}
The positive definite inner product on the positive energy subspace $\mathcal{S}^{\mathbb{C}+}$ will be given by
\begin{equation}\label{innp}
(\phi^+_1,\phi^+_2)\,=-i \omega(\overline{P^+\phi_1},P^+\phi_2)= \frac{1}{2}\left(\omega(J\phi_1,\phi_2)-i \omega(\phi_1,\phi_2)\right) \, .
\end{equation}
It is easily checked that such product is positive definite on $\mathcal{S}^{\mathbb{C}+}$ and thus the one particle Hilbert space $\mathcal{H}$ of the theory is obtained by taking the completion of $\mathcal{S}^{\mathbb{C}+}$ with respect to the above inner product.  The complex conjugate space $\bar{\mathcal{H}}$ can be thus identified with the subspace $\mathcal{S}^{\mathbb{C}-}$ and corresponds to the ``one-antiparticle" space.  The point that should be stressed (for a detailed discussion see \cite{Wald:1995yp}) is that to each choice of complex structure will correspond a inner product (and a corresponding Hilbert space construction) and vice versa.\\
It would be good at this point to make contact with the usual textbook formalism to see concrete realizations of these rather abstract constructions.  The Fourier transform of the an element $\phi\in\mathcal{S}$ can be recast as a normal mode expansion 
\begin{equation}\label{realsol2}
\phi({\bf x}, t)=\int d\mu({\bf k})\,\left [\phi^+({\bf k}) e_{{\bf k}} +\phi^-({\bf k}) \bar{e}_{{\bf k}} \right]
\end{equation}
where $e_{{\bf k}}$ is a positive energy plane wave solution 
\begin{equation}
e_{\bf k}\equiv \frac{1}{(2\pi)^{3/2}} \exp(i {\bf kx }-i\omega_{{\bf k}}t)
\end{equation}
with $\omega_{{\bf k}}=\sqrt{{\bf k}^2+m^2}$, $d\mu({\bf k})=\frac{d{\bf k}}{2\omega_{{\bf k}}}$ and the following relation between the modes (\ref{realsol2}) and the Fourier coefficients: $\phi^+({\bf k})=\tilde{\phi}(-\omega_{{\bf k}}, - {\bf k})\,,\,\, \phi^-({\bf k})= \tilde{\phi}(\omega_{{\bf k}},  {\bf k})$.  Positive and negative energy modes are defined w.r.t. the inertial time translation Killing vector $\partial_t$ and thus according to the discussion above $J=\frac{-\partial_t}{(-\partial_t\partial_t)^{1/2}}$
and in terms of the time translation generator $P_0=i\partial_t$ 
\begin{equation}
iJ=-\frac{P_0}{|P_0|}\,,\,\,\,\,\,\,\,\,\,\,\,P^{\pm}=\frac{1}{2}\left(1\pm \frac{P_0}{|P_0|}\right)\, .
\end{equation}
Using the expression for the projector above we have
\begin{equation}\label{posolcov}
\phi^+(x)=\frac{1}{(2\pi)^{3/2}}\int dk\, \delta(k^2-m^2)\, \theta(k_0)\, \tilde{\phi}(k)\exp(-ikx)=\frac{1}{(2\pi)^{3/2}}\int \frac{d{\bf k}}{2\omega_{{\bf k}}}\, \phi^+({\bf k})\exp(i {\bf kx }-i\omega_{{\bf k}}t)\, ,
\end{equation}
with $\phi^-(x)\equiv\overline{\phi^+}(x)$ and from (\ref{omega})
\begin{equation}\label{caninner}
(\phi_1^+,\phi_2^+)\equiv-i\omega(\overline{P^+\phi}_1,P^+\phi_2)=\int \frac{d{\bf k}}{2\omega_{{\bf k}}}\,\, \phi_1^-({\bf k}) \phi^+_2({\bf k})\, . 
\end{equation}
This shows how an equivalent description of the one particle Hilbert space is given by $\mathcal{H}=(M_m^+; d\mu({\bf k}))$, the space of functions on the positive mass-shell square integrable with respect to the Lorentz invariant measure $d\mu({\bf k})=\frac{d{\bf k}}{2\omega_{{\bf k}}}$.  The inner product defined above extends to a natural inner product on the whole mass-shell $M_m=M_m^+\cup M_m^-$ given by
\begin{equation}
\omega(\overline{\phi}_1,\phi_2)=i\int d^4 k\,\delta(k^2-m^2)\, \overline{\tilde{\phi}}_1(k) \tilde{\phi}_2(k)\, , 
\end{equation}
from which is is easy to write the covariant version of (\ref{caninner})
\begin{equation}\label{caninnerco}
(\phi^+_1,\phi^+_2)=\int d^4 k\,\delta(k^2-m^2)\,\theta(k^0)\, \overline{\tilde{\phi}}_1(k) \tilde{\phi}_2(k)\, . 
\end{equation}
Notice how the $\delta$-measure $d^4 k\,\delta(k^2-m^2)$ is exactly the invariant measure on the the space of functions on the homogenous space $M_m\simeq SO(3,1)/SO(3)$ we introduced in the previous section and that the complex structure, through the projection operator $P^+$, singles out a subspace of it, that of functions on the `positive energy' mass-shell.  A basis of one-particle states will be given by monochromatic plane wave solutions $e_{\bf k}$ which we denote by kets $|\bf{k}\rangle\in\mathcal{H}$.  From (\ref{posolcov}) we see that the modes associated with such solutions are
\begin{equation}
e^+_{{\bf k}}({\bf p})\equiv 2\omega_{\bf k} \,\, \delta^3({\bf p}-{\bf k})\, .
\end{equation}
It is easily checked that the normalized plane wave solutions above provide an orthogonal basis for $\mathcal{H}$ indeed 
\begin{equation}
\langle {\bf k}_1|{\bf k}_2\rangle\equiv \,(e^+_{\bf k_1},e^+_{\bf k_2})=\int \frac{d{\bf k}}{2\omega_{{\bf k}}}\,\, e_{\bf k_1}^-({\bf k}) e^+_{\bf k_2}({\bf k})=2 \omega_{\bf k_1}\delta^3({\bf k_1}-{\bf k_2})\, ,
\end{equation}
as expected.  In the rest of the paper we will show how the construction above can be extended to the quantization of a classical relativistic particle with a deformed phase space and group-valued momenta.

\section{Bending phase space}
The main point of this and the following section will be to show that when the space  $M_m$ is embedded in a group there will be quite dramatic consequences for field quantization.  In particular the introduction of curvature in momentum space leads to an ambiguity in the definition of the energy of one-particle states in terms of field modes.  This is somewhat analogous to what happens for quantum fields in curved space where one does not have a preferred notion of vacuum due to the lack of a unique way of measuring time and energy for different observers. In our case to each choice of co-ordinates on (curved) momentum space will correspond a  choice of field modes or ``linear momentum" of one-particle states.\\ 
To start off  let us make more clear the notion of ``momentum becoming group valued".  In Section II we saw how the ambient space on which the momentum sector of the phase space of a classical relativistic particle is built is the Lie algebra $\mathfrak{t}^*$ dual to the algebra of translation generators $\mathfrak{t}$.  When we say that the momentum becomes ``group valued" we mean that the Lie algebra $\mathfrak{t}^*$ acquires non-trivial Lie brackets i.e. it becomes non-abelian (unlike the case of a particle in ordinary Minkowski space).  This is to say that the dual group $T^*$ is now a non-abelian group and thus momenta, as labels of plane waves, will obey a non-abelian composition rule. Let's first see what consequences this has in general and then discuss a particular four-dimensional example.\\
First of all according to the discussion in Section II and eq. (\ref{LiePoiss}) a non-trivial Lie bracket on $\mathfrak{t}^*$ will correspond to a non-trivial Poisson-Lie structure on its dual algebra i.e. coordinate functions $x^{\mu}$ on $\mathfrak{t}$ will now have non-trivial Poisson brackets
\begin{equation}
[\cdot,\cdot]_{\mathfrak{t}^*}\neq 0\longrightarrow \{\cdot,\cdot\}_{\mathfrak{t}}\neq 0\,\, .
\end{equation}
The second consequence is that a non-trivial Lie bracket on $\mathfrak{t}^*$ induces a new structure on $\mathfrak{t}$, a ``non-trivial co-commutator" i.e. a function $\delta: \mathfrak{t}\rightarrow \mathfrak{t}\otimes \mathfrak{t}$ (which, as we will see in the next Section, will give the leading order deviation from the Leibniz rule (co-product) for a basis of the algebra of polynomials of the translation generators) defined by
\begin{equation}
\delta (Y) (\xi_1,\xi_2)\equiv \langle Y, [\xi_1,\xi_2] \rangle\,,\,\,\,\,\,\,\,\,\,\,[\cdot,\cdot]_{\mathfrak{t}^*}\neq 0\longrightarrow \delta(\cdot)_{\mathfrak{t}}\neq 0\,\, .
\end{equation}
For more details about the interplay between Poisson-Lie structures and Lie-bialgebra structures we refer the reader to \cite{Chari:1994pz}.  Notice how even when the new structures are introduced the algebra of translation generators $\mathfrak{t}$ is still abelian and thus at the  {\it  Lie algebra level} the Poincar\'e algebra is unchanged.  This means that the adjoint orbits of the Poincar\'e group on its Lie algebra are the same as in the classical case and consequently, under the dual pairing (which at the Lie algebra level does not involve any product or co-product structures) the {\it co-adjoint orbits are the same}.  This means that the classical phase space is unaffected by the introduction of a non-trivial Lie bracket on $\mathfrak{t}^*$.\\
For the case of interest to us, the $\kappa$-Poincar\'e algebra \cite{Lukierski:1992dt}, the most important new ingredient is that the dual algebra of translations gets equipped with the following bracket
\begin{equation}\label{kdual}
[P^*_{\mu},P^*_{\nu}]=-\frac{1}{\kappa} (P^*_{\mu}\delta^0_{\nu}-P^*_{\nu}\delta^0_{\mu})\, .
\end{equation}
The algebra generated by $P^*_{\mu}$ is isomorphic to the quotient Lie algebra $\mathfrak{b}\equiv\mathfrak{so}(4,1)/\mathfrak{so}(3,1)$ (see e.g. \cite{KowalskiGlikman:2004tz}).  The non-trivial co-commutators on $\mathfrak{t}$ are then given by
\begin{equation}\label{cocomm}
\delta(P^0)=0\,,\,\,\,\,\,\,\delta(P^i)=\frac{1}{\kappa} P^i \wedge P^0\, .
\end{equation}
The Lie algebra structure of $\mathfrak{t}^*=\mathfrak{b}$ will correspond to a Poisson structure on $\mathfrak{t}$ given by
\begin{equation}
\{x_{i},x_{j}\}=0\,,\,\,\,\,\{x_{0},x_{j}\}=\frac{1}{\kappa}x_j\, .
\end{equation}
Such Poisson brackets bear the same structure of the commutation relations of the so-called $\kappa$-Minkowski non-commutative space-time \cite{Majid:1994cy} but we should be careful in identifying such co-ordinates with positions of a classical relativistic particle.  Indeed as discussed in detail in Section II when building phase space from the co-adjoint orbit position variables should be constructed from the dual algebra.  As in the undeformed case we have here a choice of canonical co-ordinates on the co-adjoint orbit given by $\{p_i,x_i\}$ as discussed in section II.  In other words the classical phase space of a $\kappa$-particle is built from orbits of the undeformed Poincar\'e algebra on its dual.  Even if the latter has non trivial Lie brackets the orbits are still orbits on a linear (flat) space and thus there is no ambiguity in the choice of canonical coordinates (for more details on this conclusion drawn from an alternative approach see \cite{Arzano:2010kz}).

\section{Quantum fields and vacuum structure: a new quantization ambiguity}
As in the undeformed case plane waves will be the key ingredient in the construction the ``one-particle" Hilbert space of the theory.  In the deformed phase space setting, as remarked in the previous section, the translation group $T$ is still an abelian group and thus we can define the dual group $T^*$ as the set of plane waves (characters).  As unitary irreducible representations of $T$ we can denote plane waves as $e_{x}=\exp(i x_{\mu}P^{\mu})$ and as elements of the non-abelian group $T^*=B$, obtained by exponentiating the Lie algebra $\mathfrak{b}$ above we write $e_{p}=\exp(i p^{\mu}P^*_{\mu})$.  What is important to notice is that, unlike the undeformed case, such plane waves will have composition law w.r.t. $T$ and $T^*$ which are respectively abelian and non-abelian
\begin{equation}
e_{p}e_{q}\equiv e_{p\oplus q}\neq e_{q\oplus p}\equiv e_{q}e_{p}\, ,
\end{equation}
and
\begin{equation}
e_{x}e_{y}\equiv e_{x + y}= e_{y + x}\equiv e_{y}e_{x}\, .
\end{equation}
Likewise we will have different behaviours under group inversion 
\begin{equation}
(e_{p})^{-1}\equiv e_{\ominus p}\, ,\,\,\,\,\,\,\, (e_{x})^{-1}\equiv e_{- x}\, .
\end{equation}
The non-abelian composition rule for the $T^*$ labels can be derived in terms of the Baker-Campbell-Hausdorff formula using the Lie brackets of $\mathfrak{b}$ (see e.g. \cite{Kosinski:1999dw}). Notice however that the explicit form of such composition rule will depend on the choice of co-ordinates on the group manifold $T^*=B$. Some of this coordinate systems will correspond to group decompositions of $B$ which reflect in a splitting of the plane wave  $e_{p}$ in purely spatial and purely temporal components.  As an example we will consider the following one-parameter family of decompositions of $B$ parametrized by $0\leq|\beta|\leq 1$
\begin{equation}
e_{p}\equiv e^{-i\frac{1-\beta}{2}p^0 P^*_{0}}e^{ip^jP^*_{j}}e^{-i\frac{1+\beta}{2}p^0P^*_{0}}\ .
\end{equation}
Such parametrization will correspond to the different momentum composition rules
\begin{equation}\label{copr}
p\oplus_{\beta}q = ( p^0+ q^0;\, p^j\ e^{\frac{1-\beta}{2\kappa}q^0}+q^j\ e^{-\frac{1+\beta}{2\kappa}p^0})\, 
\end{equation}
and ``antipodes"
\begin{equation}\label{antip}
\ominus_{\beta} p = (-p^0;\, - e^{\frac{-\beta}{\kappa} p^0} p^i )\, .
\end{equation}
The non trivial behaviours of the ``deformed" plane waves above can be understood in terms of coordinate choices on the group manifold $B$.  In order to see that let us first note that as a group manifold $B$ is represented by a submanifold of de Sitter space. If we describe the latter as a four-dimensional hyper-surface embedded in five dimensional Minkowski space
\begin{equation}\label{4}
  -z_0^2 + z_1^2 + z_2^2 + z_3^2 + z_4^2 =\kappa^2\, ,
\end{equation}
it can be shown \cite{KowalskiGlikman:2004tz} that the ``momentum space" $B$ is given by the submanifold\footnote{In \cite{Freidel:2007hk} it was argued that the action of Lorentz boosts on negative frequency plane waves could take their momentum out of the submanifold describing the Lie group $B$ thus breaking Lorentz symmetry.  It was later observed by one of the authors of \cite{Freidel:2007hk}, myself and a collaborator \cite{Arzano:2009ci} that the correct way of handling the action of Lorentz generators on such antiparticle states is via their ``antipode" (see (\ref{antipode}) below).  In this way the particle/antiparticle structures and Lorentz symmetry are fully consistent.} defined by the inequality $z_0-z_4>0$.  Each choice of group splitting will correspond to a particular choice of co-ordinates on $B$ (these are obtained from acting with a matrix representation of the group element on the stability point $(0,...,\kappa)\in \mathbb{R}^{4,1} $ seen as a column vector).  For example to the ordering $\beta=1$ will correspond ``flat slicing" coordinates $p_\mu$  given by 
\begin{eqnarray}\label{bicrossp}
 {z_0}(p_0, \mathbf{p}) &=&  \kappa\sinh
{{p_0}/\kappa} + \frac{\mathbf{p}^2}{2\kappa}\,
e^{  {p_0}/\kappa}, \nonumber\\
 z_i(p_0, \mathbf{p}) &=&  - p_i \, e^{{p_0}/\kappa}, \nonumber\\
 {z_4}(p_0, \mathbf{p}) &=&  -\kappa \cosh
{{p_0}/\kappa} + \frac{\mathbf{p}^2}{2\kappa}\, e^{  {p_0}/\kappa}.
\end{eqnarray}
With a straightforward but tedious calculation one can easily obtain a general expression for co-ordinate systems associated to each value of the parameter $\beta$
\begin{eqnarray}\label{bicrosspbeta}
 {z_0}(p_0, \mathbf{p}) &= &\kappa  \left(\sinh_{+}{[p_0]} \cosh_{-}{[p_0]}+\cosh_{+}{[p_0]} \sinh_{-}{[p_0]}\right)+\nonumber\\ 
 & &+ \left(\frac{\mathbf{p}^2}{2\kappa}\right) \left(\sinh_{+}{[p_0]} \cosh_{-}{[p_0]} + \cosh_{+}{[p_0]} \cosh_{-}{[p_0]} - 
\sinh_{+}{[p_0]} \sinh_{-}{[p_0]}-\cosh_{+}{[p_0]} \cosh_{-}{[p_0]}\right) \nonumber\\
 z_i(p_0, \mathbf{p}) &=&  - p_i \, \exp_{+}{[p_0]}, \nonumber\\
 {z_4}(p_0, \mathbf{p}) &=& -\kappa  \left(\sinh_{+}{[p_0]} \cosh_{-}{[p_0]}+\cosh_{+}{[p_0]} \sinh_{-}{[p_0]}\right)+\nonumber\\ 
 & & +\left(\frac{\mathbf{p}^2}{2\kappa}\right) \left(\sinh_{+}{[p_0]} \cosh_{-}{[p_0]} + \cosh_{+}{[p_0]} \cosh_{-}{[p_0]} - 
\sinh_{+}{[p_0]} \sinh_{-}{[p_0]}-\cosh_{+}{[p_0]} \cosh_{-}{[p_0]}\right)\nonumber\\
\end{eqnarray}
where we used the compact notation $h_{\pm}[p_0]\equiv h(\frac{1\pm\beta}{2\kappa}p_0)$ for the exponential and hyperbolic functions appearing above.\\
From a mathematical point of view the different composition laws and choices of coordinates reflect the different choices of bases of the universal enveloping algebra (UEA) $U(\mathfrak{b})$ which we use to label the elements of $B$.  Recall here that roughly speaking the UEA of $\mathfrak{t}$, $U(\mathfrak{t})$, is the associative algebra of polynomials of the translation generators (see \cite{Tjin:1991me} for a pedagogical introduction).  The very important aspect of UEA of Lie algebras is that they can be endowed with an additional ``co-algebra" structure which encodes the way their representations extend to tensor product spaces.  In particular such rule of extending representations to tensor product spaces is defined by a map $\Delta: U(\mathfrak{t})\rightarrow U(\mathfrak{t})\otimes U(\mathfrak{t})$ called the ``co-product" which for ordinary UEA is nothing but the analogous of the familiar {\it Leibniz rule} for derivatives acting on products of two elements.  In mathematical language a UEA equipped with the additional co-algebra structure (and appropriate compatibility axioms) becomes a {\it Hopf algebra}.  The important thing to note is that the algebra of functions on $C^{\infty}(T^*)$ also has a natural Hopf algebra structure.  Indeed it turns out that $U(\mathfrak{t})$  is dual as a Hopf algebra to $C^{\infty}(T^*)$ and a choice of basis in $U(\mathfrak{t})$ will correspond to a choice of basis of coordinate functions on $C^{\infty}(T^*)$.  To each composition rule related to the different group splittings described above one can associate a specific co-product for the basis elements given by
\begin{equation}
\Delta(P_0)=P_0\otimes 1+1\otimes P_0\,\,\,\,\,\,\Delta(P_i)=P_i\otimes e^{\frac{1-\beta}{2\kappa}P_0}+e^{-\frac{1+\beta}{2\kappa}P_0}\otimes P_i\ , 
\end{equation}
and the corresponding antipodes, which reflect the group inversion law of $B$ on $U(\mathfrak{b})$, given by
\begin{equation}\label{antipode}
S(P_0)=-P_0\,\,\,\,\,\, S(P_i)= -e^{\frac{\beta}{\kappa}P_0} P_i\ .
\end{equation}
From these basic ingredients, under certain compatibility requirements for the action of the Lorentz group on the deformed momentum space, one can reconstruct the structure of the whole deformed $\kappa$-Poincar\'e algebra (see \cite{Majid:1994cy} for details of the construction and \cite{Arzano:2007qp} for a condensed review of the $\kappa$-Poincar\'e algebra).  Notice that the $1/\kappa$ term of the antisymmetric part of the different co-products which reproduces the co-commutator (\ref{cocomm}) {\it does not} depend on the choice of co-ordinates and thus all the structures at the level of Lie algebra are {\it uniquely} defined which means that there is no ambiguity in describing the phase space of a classical relativistic particle even when the deformations are introduced. \\
After this digression on the structure of the dual group $T^*=B$ we turn back to our main task which is the definition of a one-particle Hilbert space from the classical phase space described in the previous section.  Now that we have identified the (deformed) space of characters, in analogy with the undeformed case, we will consider the orbits under the action of the Lorentz group.  Indeed on elements of $T^*=B$ one can define a natural action\footnote{Recall even if the action of the Lorentz group on $B$ is not a representation the action on the space of functions on $B$ does provide a representation.} which is induced from the action of the Lorentz group on the five dimensional Minkowski space, in which the de Sitter hyperboloid is embedded, keeping the $z_4$ co-ordinate fixed.  
This will lead to an action of the usual Lorentz group $SO(3,1)$ leaving invariant the hyperboloid \cite{Arzano:2009ci}
\begin{equation}\label{4}
   -z_0^2 + z_1^2 + z_2^2 + z_3^2  =\kappa^2-\tilde{m}^2\, ,
\end{equation}
which describes the ``deformed" mass-shell given by
\begin{equation}
M^{\kappa}_m \equiv\{\gamma e_{p}: e_{p}\in B,\, \gamma\in SO(3,1) \}\, .
\end{equation}
As for the undeformed mass shell described in Section II, the space $M^{\kappa}_m $ as the orbit of a symmetry group will have a natural geometrical interpretation as a homogenous space.  The deformed one-particle Hilbert space will be built from the space of functions on such homogenous space $C^{\infty}(M^{\kappa}_m)$.  As discussed above a choice of co-ordinates on $B$ is associated to a choice of basis of $U(\mathfrak{b})$ and to the hyperboloid above will correspond with an invariant mass Casimir operator $C_1(P)\in U(\mathfrak{b})$.  Functions on the mass-shell $\phi \in C^{\infty}(M^{\kappa}_m)$ will thus satisfy the ``wave equation"
\begin{equation}\label{waveq}
C_1(P)\,\phi =m^2 \phi \,, 
\end{equation}
where $m^2=\tilde{m}^2-\kappa^2$.  In particular (\ref{waveq}) will hold for plane waves themselves.  Notice that for any Lie group $G$ the space of complex valued functions square integrable w.r.t. the inner product defined using the Haar measure $d\mu (g)$
\begin{equation}
(f_1,f_2)=\int_G d\mu (g)\, \,  \bar{f}_1(g)\, f_2(g)\, .
\end{equation}
defines a Hilbert space.  In our case, as functions on a homogeneous space we can define a natural invariant measure and a inner product on $C^{\infty}(M^{\kappa}_m)$ (see discussion in Section II) with the latter given by 
\begin{equation}\label{kinner}
(\phi_1,\phi_2)_{\kappa}=\int_B d\mu (p)\, \, \delta(\mathcal{C}_1(p))\,\, \bar{\phi}_1(p)\, \phi_2(p)\, .
\end{equation}
Here $d\mu (p)$ is the left-invariant Haar measure on $T^*=B$ \cite{Freidel:2007hk} which in cartesian and flat slicing co-ordinates reads respectively
\begin{equation}\label{19a}
d\mu \equiv \frac{1}{(2\pi)^4\,z_4}\, d z_0\, d^3\mathbf{z} = \frac{e^{3p_0/\kappa}}{(2\pi)^4}\, dp_0\, d^3\mathbf{p}\,.
\end{equation}
To define a Hilbert space from $C^{\infty}(M_m)$ we need to find a criterion which ensures that the inner product (\ref{kinner}) is positive definite. As discussed at length in Section III this entails the introduction of a complex structure on $C^{\infty}(M_m)$.  Roughly speaking this corresponds to a choice of a ``time-like" element of $P_0 \in U(\mathfrak{b})$ such that
\begin{equation}
P_0 \, \phi(p)^{\pm}=  \omega^{\pm}(p)\, \phi(p)^{\pm}\,, 
\end{equation}
i.e. the equivalent of an energy co-ordinate function on the homogenous space $M^{\kappa}_m$.  The complex structure will be, as usual, given by
\begin{equation}\label{complexs}
J=i \frac{P_0}{|P_0|}\, ,
\end{equation}
(properly speaking such element is not in the UEA but in the ``enveloping field" \cite{Barut:1986dd}) and, as in the undeformed case, can be used to define positive and negative energy projection operators.\\  
Now we come to our main point.  In order to choose the energy operator $P_0$ from which we define the complex structure we need to make an explicit choice of basis in the commutative UEA $U(\mathfrak{t})$ with which we decompose the element $C_1(P)$.  In ordinary local QFT the requirement of ``local action" of a symmetry generator singles out a {\it unique choice} of basis of translation generators $P_0,\,P_i$ for which $C_1(P)=P_0^2-{\bf P}^2_i$.  Indeed in this case a choice of cartesian co-ordinates on $C^{\infty}(\mathbb{R}^{3,1})$ will correspond to the set of basis elements $P_0,\,P_i$  of $U(\mathbb{R}^{3,1})$ for which
\begin{equation}
\Delta P_{\mu} = P_{\mu} \otimes 1 + 1 \otimes P_{\mu}\, .
\end{equation}
Elements of a UEA for which the co-product has such form are called ``primitive".  In everyday language the trivial form of the co-product above is telling us that primitive elements act according to the Leibnitz rule i.e. additively and thus are ``local symmetry generators" (see \cite{Arzano:2007nx} for a detailed discussion).\\ 
In our deformed setting the peculiarity of $U(\mathfrak{b})$ is that now {\it there is no choice} of a commuting set of primitive elements with which we decompose the Casimir.  Indeed since the dual Hopf algebra of $U(\mathfrak{b})$ is, loosely speaking, the algebra of functions on the {\it non-abelian} group $B$ the co-product of $U(\mathfrak{b})$ {\it no matter which basis we choose} will be non co-commutative, namely $\sigma \circ \Delta \neq \Delta$ (where $\sigma (a\otimes b) = b \otimes a)$.  In other words the action of translation generators will be non-Leibniz and non-symmetric for ANY choice of basis of $U(\mathfrak{b})$. This is the most profound and truly ``basis independent" statement in the context of deformed relativistic symmetries.  We thus conclude that there is no preferred choice of translation symmetry generators from which we can define an energy coordinate function on $M^{\kappa}_m$ and thus {\it no preferred choice of complex structure} in constructing the one-particle Hilbert space of a relativistic particle with curved momentum space.\\
Note that in QFT in curved space one faces an analogous situation: in this case the ambiguity in the the definition of the complex structure $J$ comes from the fact that there is no global time-like Killing vector that can be used to define such object.  In the most optimistic cases one has a preferred notion of the time-translation only in certain regions of space-time and this ultimately leads to particle production when one evolves from a region to another.  Notice that while for us to allow for a generalization of the quantization formalism to curved momentum space we had to start with a phase space described in terms of co-adjoint orbits in QFT in curved space the starting point is the phase space of the field described as solutions of the equation of motion which is well defined on any global hyperbolic manifold (which can have no global symmetries at all).

\section{Field modes and vacuum fluctuations}
We now give a concrete realization of the deformed one-particle Hilbert space and introduce tools to describe the behaviour of deformed field modes.  Let us focus on the choice of basis $P_0, P_i$ in $U(\mathfrak{b})$ related to the ``flat slicing" co-ordinates (\ref{bicrossp}) i.e. to the group splitting parameter $\beta=1$.  The wave equation defining the mass-shell is given by the mass Casimir which for such choice of basis reads 
\begin{equation}
\mathcal{C}_{\kappa}(P)=\left(2\kappa \sinh\left(\frac{P_0}{2\kappa}\right)\right)^2-{\bf P}^2 e^{P_0/\kappa}\,.
\end{equation}
For simplicity we focus on the massless case.  For on-shell plane waves $e_{\bf p}\equiv \{e_p:  \mathcal{C}_{\kappa}(P)\, e_p=0\}$, and in general of any function on $M^{\kappa}_{m}$, the generator $P_0$ will read off the energy coordinate 
\begin{equation}
P_0\,\, e_{\bf p} = \omega^{\pm}_{\kappa}({\bf p})\,\, e_{\bf p}\, ,
\end{equation}
with
\begin{equation}
\omega^{\pm}_{\kappa}({\bf p})=-\kappa \log\left(1\mp \frac{|{\bf p}|}{\kappa}\right) \, .
\end{equation}
We can now use $P_0$ to define the complex structure (\ref{complexs}) and the operator $P^+= 1/2 1- iJ$ to project a generic element of $C^{\infty}(M^{\kappa}_m)$ on the positive energy subspace  $C^{\infty}(M^{\kappa +}_m)$.  The inner product on such space given by 
\begin{equation}
(\phi_1,\phi_2)_{\kappa}=\int_{M^{\kappa +}_m} \frac{d\mu({\bf p})}{2\omega_{\kappa}({\bf p})}\,\, \bar{\phi}_1({\bf p})\, \phi_2({\bf p})\, ,
\end{equation} 
(we omitted for the $+$ superscripts for notational clarity), which can be written in covariant form as \cite{Arzano:2007ef}
\begin{equation}\label{sinner}
(\phi_1,\phi_2)_{\kappa}=\int_{B} d\mu(p)\,\,\delta(\mathcal{C}_1(p))\, \theta(p_0)\, \overline{\tilde{\phi}}_1(p)\, \tilde{\phi}_2(p)\, ,
\end{equation}
is indeed positive definite and thus turns $C^{\infty}(M^{\kappa +}_m)$ into our deformed one-particle Hilbert space $\mathcal{H}_{\kappa}$.  Using the group Fourier transform discussed in Section II we can write the ``space-time" counterpart of  $ \phi({\bf p})\in C^{\infty}(M^{\kappa +}_m)$
\begin{equation}
\phi(x) = \int_{B} d\mu(p)\,\,\delta(\mathcal{C}_1(p))\, \theta(p_0)\, \tilde{\phi}(p)\, e_p(x) = \int_{M^{\kappa +}_m} \frac{d\mu({\bf p})}{2\omega_{\kappa}({\bf p})}\,\,\, \phi({\bf p})\, e_{{\bf p}}(x)
\end{equation}
which shows how, due to the group nature of the plane waves $e_p(x)$, the fields $\phi(x)$ form a non-commutative algebra and thus the Fourier transformed version of elements of $\mathcal{H}_{\kappa}$ describe the one-particle Hilbert space of a {\it non-commutative quantum field theory}.\\
For a practical description of the states $\mathcal{H}_{\kappa}$ we can introduce a normalized basis of delta functions\footnote{Recall that the Dirac delta for functions on a group $G$ is such that 
\begin{equation*}
\int_G d\mu(g) \delta(g) f(g) = f(e)\,,\,\,\,\,\,\,\,\, \int_G d\mu(g) \delta(gh^{-1}) f(g)=f(h)\, ,
\end{equation*}
where $g,h \in G$ and $e$ is the unit element which in the notation used in the preceding Sections this reads
\begin{equation*}
\int_B d\mu(p) \delta(p) f(p) = f(0)\,,\,\,\,\,\,\,\,\, \int_B d\mu(p) \delta(p\oplus(\ominus q)) f(p)=f(q)\, .
\end{equation*}} which correspond to the ``modes'' of the on-shell plane waves $e_{\bf p}$
\begin{equation}
e_{{\bf p}}({\bf k})\equiv 2\omega_{\kappa}({\bf k}) \,\, \delta^3({\bf p}\oplus (\ominus {\bf k}))\, ,
\end{equation}
where $\oplus$ and $\ominus$ denote respectively the (non-abelian) composition and antipode for spatial momenta which can be read off (\ref{copr}) and (\ref{antip}) and are explicitely given by
\begin{equation}
{\bf p}\oplus {\bf q} = {\bf p} + e^{\frac{-p^0}{\kappa}} {\bf q} \,,\,\,\,\,\,\,\,\,\,\,\,\,\,\, \ominus {\bf p} = - e^{\frac{p^0}{\kappa} } {\bf p}\, .
\end{equation}
Introducing a bra-ket notation $e_{\bf p} \equiv |{\bf p} \rangle$ we have for the inner product of one-particle states \cite{Arzano:2007ef}
 \begin{equation}
\langle{\bf k_1}|{\bf k_2} \rangle \equiv (e_{\bf k_1},e_{\bf k_2})_{\kappa}=2\omega_{\kappa}({\bf k}_1) \,\, \delta^3({\bf k}_1\oplus (\ominus {\bf k}_2))\,.
\end{equation}
Of course the same construction above can be repeated for any other choice of the group splitting parameter $\beta$.  In this case the Hilbert space $\mathcal{H}^{\kappa}_{\beta}$ will be spanned by basis vectors $|{\bf k} \rangle_{\beta}$ bearing a different relation between energy and linear momentum through $\omega^{\beta}_{\kappa}({\bf k})$ and a different composition rule for the eigenvalues of the deformed translation generators $P^{\beta}_{\mu}$.  Notice also that unlike the case of quantum fields in curved space the different Hilbert space constructions share the {\it same} vacuum state.\\  
Within the context of one-particle quantization we can proceed a step further and study the basic observables of the theory in order to get some insight on the vacuum structure and quantum fluctuations of the theory.  One-particle observables will be given by the quantized counterpart of classical observables i.e. functions on phase space.  The latter can be written in terms of the symplectic structure as $\mathcal{O}_{\phi}\equiv \omega(\phi, \cdot)$ with $\phi\in C^{\infty}(M^{\kappa}_m)$.  Quantization of such observable gives the most general expression of the field operator $\hat{\mathcal{O}}_{\phi}\equiv\Psi(\phi)$ which for specific choices of $\phi$ reduces to the familiar field operator (see \cite{Bongaarts:1971cu} for a nice discussion).  The one-particle creation and annihilation operator will be obtained upon quantization of the following functions on phase space
\begin{eqnarray}
a(\phi)(\cdot)&\equiv\frac{1}{2}\left(\omega(J\phi,\cdot)-i\omega(\phi,\cdot)\right)&=\,\langle\phi,\cdot \rangle\\\, ,
a^*(\phi)(\cdot)&\equiv\frac{1}{2}\left(\omega(J\phi,\cdot)+i\omega(\phi,\cdot)\right)&=\,\langle\cdot,\phi \rangle\, .
\end{eqnarray} 
In terms of the delta function basis written above we denote the quantized counterparts of such functions by
\begin{equation}
a(\bar{e}_{\bf k})\equiv a({\bf k}) \, ,\,\,\,\,\,\,\,\, a^{\dagger}(e_{\bf k})\equiv a^{\dagger}({\bf k})\, ,
\end{equation}
so that
\begin{equation}\label{aphi}
a(\phi)= \int \frac{d\mu({\bf k})}{2\omega_{\kappa}({\bf k})}\, \phi({\bf k})\, a({\bf k})\,.
\end{equation}
and 
\begin{equation}\label{aphi+}
a^{\dagger}(\phi)= \int \frac{d\mu({\bf k})}{2\omega_{\kappa}({\bf k})}\, \phi(\ominus{\bf k})\, a^{\dagger}({\bf k})\, ,
\end{equation}
where the antipode in the last expression comes from the reality condition on the classical phase space element $\phi\in C^{\infty}(M^{\kappa}_m)$.   The ``generalized" filed operator can be written in terms of such creation and annihilation operators\footnote{Let us remark here that, as widely discussed in the literature \cite{Arzano:2007ef, Arzano:2008bt, Bu:2006dm, Govindarajan:2008qa, Young:2008zm}, the extension of the creation and annihilation operators defined above to the multi-particle sector of the theory is highly non-trivial.  In fact in the construction of a deformed Fock space the non-symmetric nature of the co-product requires a ``momentum-shifting" symmetrization \cite{Arzano:2007ef, Arzano:2009wp}. The existence of a covariant deformed symmetrization procedure depends on the availability of an operator known as quantum $R$-matrix (see \cite{ Arzano:2008bt, Young:2008zm} for an extended discussion) whose explicit construction for the $\kappa$-Poincar\'e algebra has been a topic of various studies without a commonly agreed outcome.  We should notice however that our analysis goes beyond the illustrative example of $\kappa$-deformation and, for example, would also apply to the case of deformed relativistic symmetries described by the so-called Lorentz double \cite{Bais:2002ye}.  For such models one has a rather straightforward definition of $R$-matrix and thus, in principle, no obstacles in the construction of a consistent Fock space.} as
\begin{equation}
\Psi(\phi)=i(a(\phi)-a^{\dagger}(\phi))\,,
\end{equation}
and from $\Psi(\phi)$ we can write down the field mode operator or the quantum equivalent of the classical oscillator coordinate.  Indeed using the expansions (\ref{aphi}) and (\ref{aphi+})
\begin{equation}
\Psi(\phi)=i \int \frac{d\mu({\bf k})}{2\omega_{\kappa}({\bf k})} \tilde{\phi}({\bf k}) \left(a({\bf k}) + \mathcal{J}_{\ominus}({\bf k})\, a^{\dagger}(\ominus{\bf k})\right)\, ,
\end{equation}
with $\mathcal{J}_{\ominus}({\bf k})$ defined by $d\mu(\ominus {\bf k})= \mathcal{J}_{\ominus}({\bf k})d\mu({\bf k})$.  We have for the Schroedinger picture field mode operator
\begin{equation}
\hat{\varphi}_{\kappa}({\bf k})\equiv\frac{1}{2\omega_{\kappa}({\bf k})}  (a({\bf k}) + \mathcal{J}_{\ominus}({\bf k})\, a^{\dagger}(\ominus{\bf k}))\,.
\end{equation}
We can evolve $\hat{\varphi}_{\kappa}({\bf k})$ in time using the translation generator $P_0$ obtaining the field mode operator in the Heisenberg representation 
\begin{equation}
\hat{\varphi}_{\kappa}({\bf k},t)\equiv\frac{1}{2\omega_{\kappa}({\bf k})}  (a({\bf k}) \exp(-i \omega_{\kappa}({\bf k})t)+ \mathcal{J}_{\ominus}({\bf k})\, a^{\dagger}(\ominus{\bf k})\exp(i \omega_{\kappa}({\bf k})t))\,.
\end{equation}
We can now take the expectation value of the product of two mode-field operators above in the vacuum state $|0\rangle$ such that $a^{\dagger}({\bf k})|0\rangle\equiv |{\bf k}\rangle$ and $a({\bf k})|0\rangle\equiv 0\,\,\, \forall\,{\bf k}$. Thus we obtain the deformed equivalent of the spatial Fourier transform of the two-point function 
\begin{equation}
G_+({\bf k}_1,t;{\bf k}_2,s)\equiv\langle 0| \hat{\varphi}_{\kappa}({\bf k}_1,t)  \hat{\varphi}_{\kappa}({\bf k}_2, s) |0\rangle =\frac{\delta^3({\bf k}_1\oplus{\bf k}_2)}{2\omega_{\kappa}({\bf k}_1)} \mathcal{J}_{\ominus}({\bf k}_1)\exp(-i\omega_{\kappa}({\bf k}_1)(t-s)) \, .
\end{equation}
This provides us with the fundamental building block for $\kappa$-deformed field theory and for all the applications in which the two-mode point function plays a fundamental role.\\ 
As an immediate application of the formalism introduced we can calculate the {\it vacuum fluctuations} of the field modes $\hat{\varphi}_{\kappa}({\bf k})$ which will be given by
\begin{equation}
\delta\hat{\varphi}_{\kappa}({\bf k})=(\langle 0| \hat{\varphi}_{\kappa}({\bf k}) \hat{\varphi}^{\dagger}_{\kappa}({\bf k})|0\rangle)^{1/2}\sim \frac{\mathcal{J}_{\ominus}({\bf k})}{2\omega_{\kappa}({\bf k})} \, .
\end{equation}
For the illustrative case of $\beta=1$ we have that $\mathcal{J}_{\ominus}({\bf k})=\exp(-3\omega_{\kappa}({\bf k})/\kappa)$ and thus
\begin{equation}
\delta\hat{\varphi}_{\kappa}({\bf k})\rightarrow 0\, ,\,\,\,\,\,\,\,\,\ |{\bf k}|\rightarrow \kappa\, ,
\end{equation}
i.e. quantum fluctuations {\it freeze} when the modulus of the linear momentum of the field mode approaches the value of the deformation parameter $\kappa$.  Notice how such result heavily relies on the definition of linear modes for the field one is choosing.  From this point of view the study of mode fluctuations seem to be a good candidate to establish, via some physical requirement, whether or not a ``preferred" notion of field mode exist in the quantization procedure we outlined.  Such question will be addressed in future work.

\section{Summary}
We presented a detailed account of the quantization of a relativistic particle with momentum space given by a group manifold.  This was done starting from a description of the phase space of the particle as a co-adjoint orbit of the relativistic symmetry group.  The reason for adopting such formulation was twofold: on one side it is naturally connected  with the description of the corresponding classical and quantum field theory spaces of states on the other hand allows for generalizations to models of relativistic particles with group valued momenta for which a notion of configuration space is less straightforward. We discussed how, in general, at the phase space level ``curving" momentum space boils down to the introduction of a non-trivial Lie bracket on the dual Lie algebra of translations. In particular we considered the ``group" momentum space associated with $\kappa$-deformations of the Poincar\'e algebra which is obtained by exponentiating the $\kappa$-Minkowski Lie brackets and which, as a manifold, is given by a sub-manifold of de Sitter space.  Our analysis shows that, at least at the kinematical level, there is no effect of such deformations on the classical phase space of a single relativistic particle, a result which confirms what suggested in \cite{Arzano:2010kz}.\\
Effects of the deformation do indeed appear, and quite dramatically, at the quantum level. We recalled how a necessary step in the construction of a quantum Hilbert space from a classical field's phase space is the introduction of a {\it complex structure} which defines the notion of positive and negative energy states.  We showed that, for a deformed field theory related to a relativistic particle with curved momentum space, this step is non-trivial since it involves a choice of basis in the algebra of polynomials of the generators of deformed translations.  As for field quantization in curved space-time, in a deformed setting one does not have a criterion to pick a preferred notion of energy (and linear momentum).  This is to contrast with ordinary local quantum field theory in which such criterion exists and consists in picking a basis of translation generators which act according to the Leibniz rule on tensor product states i.e. whose momenta combine according to usual addition.  Even though our discussion was limited to the example of $\kappa$-deformed momentum space, the conclusion we reach applies to {\it any} field theory with group valued momenta and in particular to the ``quantum double" of the Lorentz group, a deformation of the Poincar\'e algebra relevant for relativistic particles coupled to three dimensional gravity \cite{Bais:2002ye}.\\
The tools introduced in the discussion of the quantization of the $\kappa$-deformed field theory were used in the last section to provide a concrete realization of a $\kappa$-one-particle Hilbert space.  We defined the basic field observable of the theory and were able to explicitly derive the quantized mode operators which were used to write down the deformed two-point function in the linear momentum representation and the vacuum fluctuations of the modes, which, as expected, exhibit a non-trivial behaviour when their modulus gets closer to the (UV) deformation scale $\kappa$.  This further step in the understanding the quantum properties of $\kappa$-deformed field theories finally opens the window to what we think are most promising applications of these models namely their use for investigating trans-planckian issues \cite{Corley:1996ar, Martin:2000xs, Parentani:2007dw}  in semi-classical gravity from cosmology to black hole radiance. 
 
\begin{acknowledgments}
The author would like to thank Jerzy Kowalski-Glikman for many valuable discussions and encouragement and Tomasz Trzasniewski for comments on the last version of the paper.
This research is supported by the EC's Marie Curie Actions. I would like to thank the Perimeter Institute for Theoretical Physics for hospitality while part of this project was being carried out.  Research at Perimeter Institute for Theoretical Physics is supported in part by the Government of Canada through NSERC and by the Province of Ontario through MRI.
\end{acknowledgments}

\end{document}